%% file: main.tex
\documentclass{article}

\usepackage{microtype}
\usepackage{graphicx}
\usepackage{subfigure}
\usepackage{booktabs} 

\usepackage{hyperref}

\usepackage[accepted]{icml2024}

\usepackage{amsmath}
\usepackage{amssymb}
\usepackage{mathtools}
\usepackage{amsthm}
\usepackage{multirow}
\usepackage{booktabs}
\usepackage[capitalize,noabbrev]{cleveref}

\theoremstyle{plain}

\theoremstyle{definition}

\theoremstyle{remark}

\usepackage{soul} 
\usepackage{color, xcolor} 
\usepackage[textsize=tiny]{todonotes}
\definecolor{myred}{RGB}{247,206,205} 
\definecolor{mygreen}{RGB}{204,204,251} 

\icmltitlerunning{Submission and Formatting Instructions for ICML 2024}

\begin{document}

\twocolumn[
\icmltitle{IIANet: An Intra- and Inter-Modality Attention Network for \\ Audio-Visual Speech Separation}




\begin{icmlauthorlist}
\icmlauthor{Kai Li}{yyy}
\icmlauthor{Runxuan Yang}{yyy}
\icmlauthor{Fuchun Sun}{yyy}
\icmlauthor{Xiaolin Hu}{yyy,comp,sch}
\end{icmlauthorlist}

\icmlaffiliation{yyy}{Department of Computer Science and Technology, Institute for AI, BNRist, Tsinghua University, Beijing 100084, China}
\icmlaffiliation{comp}{Tsinghua Laboratory of Brain and Intelligence (THBI), IDG/McGovern Institute for Brain Research, Tsinghua University, Beijing 100084, China}
\icmlaffiliation{sch}{Chinese Institute for Brain Research (CIBR), Beijing 100010, China}

\icmlcorrespondingauthor{Xiaolin Hu}{xlhu@tsinghua.edu.cn}

\icmlkeywords{Machine Learning, ICML}

\vskip 0.3in
]




\begin{abstract}
\input{abstract}
\end{abstract}

\section{Introduction}
\input{introduction}

\section{Related work}

\input{related_work}

\section{The proposed model}
\input{method}

\section{Experiments}
\input{experiments}

\section{Conclusion and discussion}
\input{conclusion}

\section{Impact statements}
This paper presents work whose goal is to advance the field of Machine Learning. There are many potential societal consequences of our work, none which we feel must be specifically highlighted here.

\bibliography{example_paper}
\bibliographystyle{icml2024}

\newpage
\appendix
\onecolumn
\input{appendix}

\end{document}

%% file: abstract.tex
Recent research has made significant progress in designing fusion modules for audio-visual speech separation. However, they predominantly focus on multi-modal fusion at a single temporal scale of auditory and visual features without employing selective attention mechanisms, which is in sharp contrast with the brain. To address this issue,  We propose a novel model called \textit{Intra- and Inter-Attention Network (IIANet)}, which leverages the attention mechanism for efficient audio-visual feature fusion. IIANet consists of two types of attention blocks: intra-attention (IntraA) and inter-attention (InterA) blocks, where the InterA blocks are distributed at the top, middle and bottom of IIANet. 
Heavily inspired by the way how human brain selectively focuses on relevant content at various temporal scales, these blocks maintain the ability to learn modality-specific features and enable the extraction of different semantics from audio-visual features. Comprehensive experiments on three standard audio-visual separation benchmarks (LRS2, LRS3, and VoxCeleb2) demonstrate the effectiveness of IIANet, outperforming previous state-of-the-art methods while maintaining comparable inference time. In particular, the fast version of IIANet (\textit{IIANet-fast}) has only 7\% of CTCNet’s MACs and is 40\% faster than CTCNet on CPUs while achieving better separation quality, showing the great potential of attention mechanism for efficient and effective multimodal fusion.

%% file: introduction.tex
In our daily lives, audio and visual signals are the primary means of information transmission, providing rich cues for humans to obtain valuable information in noisy environments \cite{cherry1953some, arons1992review}. This innate ability is known as the ``cocktail party effect", also referred to as ``speech separation" in computer science. Speech separation possesses extensive research significance, such as assisting individuals with hearing impairments \cite{summerfield1992lipreading}, enhancing the auditory experience of wearable devices \cite{ryumin2023audio}, etc.

Researchers have previously sought to address the cocktail party effect through audio streaming alone \cite{luo2019conv, luo2020dual, hu2021speech}. This task is called audio-only speech separation (AOSS). However, the quality of the separation system may decline dramatically when speech is corrupted by noise \cite{afouras2018deep}. To address this issue, many researchers have focused on audio-visual speech separation (AVSS), achieving remarkable progress. Existing AVSS methods can be broadly classified into two categories based on their fusion modules: CNN-based and Transformer-based. CNN-based methods \cite{gao2021visualvoice, li2022audio, wu2019time} exhibit lower computational complexity and excellent local feature extraction capabilities, enabling the extraction of audio-visual information at multiple scales to capture local contextual relationships. Transformer-based methods \cite{lee2021looking, rahimi2022reading, montesinos2022vovit} can leverage cross-attention mechanisms to learn associations across different time steps, effectively handling dynamic relationships between audio-visual information. 

By comparing the working mechanisms of these AVSS methods and the brain, we find that there are two major differences. First, auditory and visual information undergoes neural integration at multiple levels of the auditory and visual pathways, including the thalamus \cite{halverson2006medial,cai2019distinct}, A1 and V1 \cite{mesik2015functional,eckert2008cross}, and occipital cortex \cite{ghazanfar2006neocortex,stein2008multisensory}. However, most AVSS methods integrate audio-visual information only at either coarse \cite{gao2021visualvoice, lee2021looking} or fine \cite{li2022audio, montesinos2022vovit, wu2019time, rahimi2022reading} temporal scale, thus ignoring semantic associations across different scales. 

Second, numerous evidences have indicated that the human brain employs selective attention to focus on relevant speech in a noisy environment \cite{cherry1953some, golumbic2013mechanisms}. This also modulates neural responses in a frequency-specific manner, enabling the tracked speech to be distinguished and separated from competing speech \cite{golumbic2013mechanisms, mesgarani2012selective, o2015attentional, kaya2017modelling, cooke2001auditory,aytar2016soundnet}. The Transformer-based AVSS methods adopt attention mechanisms \cite{rahimi2022reading, montesinos2022vovit}, but they commonly employ the Transformer as the backbone network to extract and fuse audio-visual features, and the attention is not explicitly designed for selecting speakers. 
In contrast to above complex fusion strategies, our approach aims to design a simple and efficient fusion module based on the selective attention mechanism. 

To achieve this goal, inspired by the structure and function of the brain, we propose a  concise and efficient audio-visual fusion scheme, called Intra- and Inter-Attention Network (IIANet), which leverages different forms of attention\footnote{Please note that here the intra-attention and inter-attention are irrelevant to Transformer.} to extract diverse semantics from audio-visual features. IIANet consists of two hierarchical unimodal networks, mimicking the audio and visual pathways in the brain, and employs two types of attention mechanisms: intra-attention (IntraA) within single modalities and inter-attention (InterA) between modalities. 

Following a recent AOSS model \cite{li2022efficient}, the IntraA has two forms of top-down attention, global and local, which aim to enhance the ability of the model to select relevant information in unimodal networks with the guidance of higher-level features. This mechanism is inspired by the massive top-down neural projections discovered in both auditory \cite{guinan2006olivocochlear,budinger2008non} and visual \cite{angelucci2002anatomical,felleman1991distributed} pathways. The InterA mechanism aims to select relevant information in one modality with the help of the features in the other unimodal network. It is used at all levels of the unimodal networks. Roughly speaking, the InterA block at the top level (InterA-T) corresponds to higher associate cortical areas such as the frontal cortex and occipital cortex \cite{raij2000audiovisual, keil2012variability, stein2008multisensory}, and at the bottom level (InterA-B) corresponds to the thalamus \cite{halverson2006medial, cai2019distinct}. The InterA blocks at the middle levels (InterA-M) reflect the direct neural projections between different auditory areas and visual areas, e.g., from the core and belt regions of the auditory cortex to the V1 area  \cite{falchier2002anatomical}.


On three AVSS benchmarks, we found that IIANet surpassed the previous SOTA method CTCNet \cite{li2022audio} by a considerable margin, while the model inference time was nearly on par with CTCNet. We also built a fast version of IIANet, termed IIANet-fast. It achieved an inference speed on CPU that was more than twice as fast as CTCNet and still yielded better separation quality on three benchmark datasets.

%% file: related_work.tex
\subsection{Audio-visual speech separation}
Incorporating different modalities aligns better with the brain's processing \cite{calvert2001crossmodal, bar2007proactive}. Some methods \cite{gao2021visualvoice, li2022audio, wu2019time, lee2021looking, rahimi2022reading, montesinos2022vovit} have attempted to add visual cues in the AOSS task to improve the clarity of separated audios, called audio-visual speech separation (AVSS). They have demonstrated impressive separation results in more complex acoustic environments, but these methods only focus on audio-visual features fusion at a single temporal scale in both modality, e.g., only at the finest \cite{wu2019time, li2022audio} or coarsest \cite{gao2021visualvoice, lee2021looking} scales, which restrains their efficient utilization of different modalities' information. While CTCNet's fusion module \cite{li2022audio} attempts to improve separation quality by utilizing visual and auditory features at different scales, it still focuses on fusion at the finest temporal scales. This unintentionally hinders the effectiveness of using visual information for guidance during the separation process. Moreover, some of the AVSS models \cite{lee2021looking, montesinos2022vovit} employ attention mechanisms within the process of inter-modal fusion, while the significance of intra-modal attention mechanisms is overlooked. This deficiency constrains their separation capabilities, as the global information within the auditory modality plays a crucial role in enhancing the clarity of the separated audio \cite{li2022efficient}. Notably, the fusion module of these methods \cite{lee2021looking, montesinos2022vovit} tend to use a similarity matrix to compute the attention, which dramatically increases the computational complexity. In contrast, our proposed IIANet uses the sigmoid function and element-wise product for selective attention, which can efficiently integrate audio-visual features at different temporal scales.

\subsection{Attention mechanism in speech separation}
Attention is a critical function of the brain \cite{rensink2000dynamic,corbetta2002control}, serving as a pivotal mechanism for humans to cope with complex internal and external environments. Our brain has the ability to focus its superior resources to selectively process task-relevant information \cite{schneider2013selective}. It is widely accepted that numerous brain regions are involved in the transmission of attention, forming an intricate neural network of interconnections \cite{posner1990attention,fukushima1986neural}. These regions play distinct roles, enabling humans to selectively focus on processing certain information at necessary times and locations while ignoring other perceivable information. While this ability is known to work well along with sensory or motor stimuli, it also enables our brain to maintain the ability to process information in real-time even in the absence of these stimuli \cite{tallon2012neural,koch2007attention}. 

Recently, \citet{kuo2022inferring} constructed artificial neural networks to simulate the attention mechanism of auditory features, confirming that top-down attention plays a critical role in addressing the cocktail party problem. Several recent AOSS models \cite{shi2018listen,li2022efficient,chen2023neural} have utilized top-down attention in designing speech separation models. These approaches were able to reduce computational costs while maintaining audio separation quality. They have also shown excellent performance dealing with complex mixture audio. However, the efficient integration of top-down and lateral attention between modalities in AVSS models remains an unclear aspect.


%% file: method.tex

\begin{figure*}[t]
\centering
\includegraphics[width=0.8\linewidth]{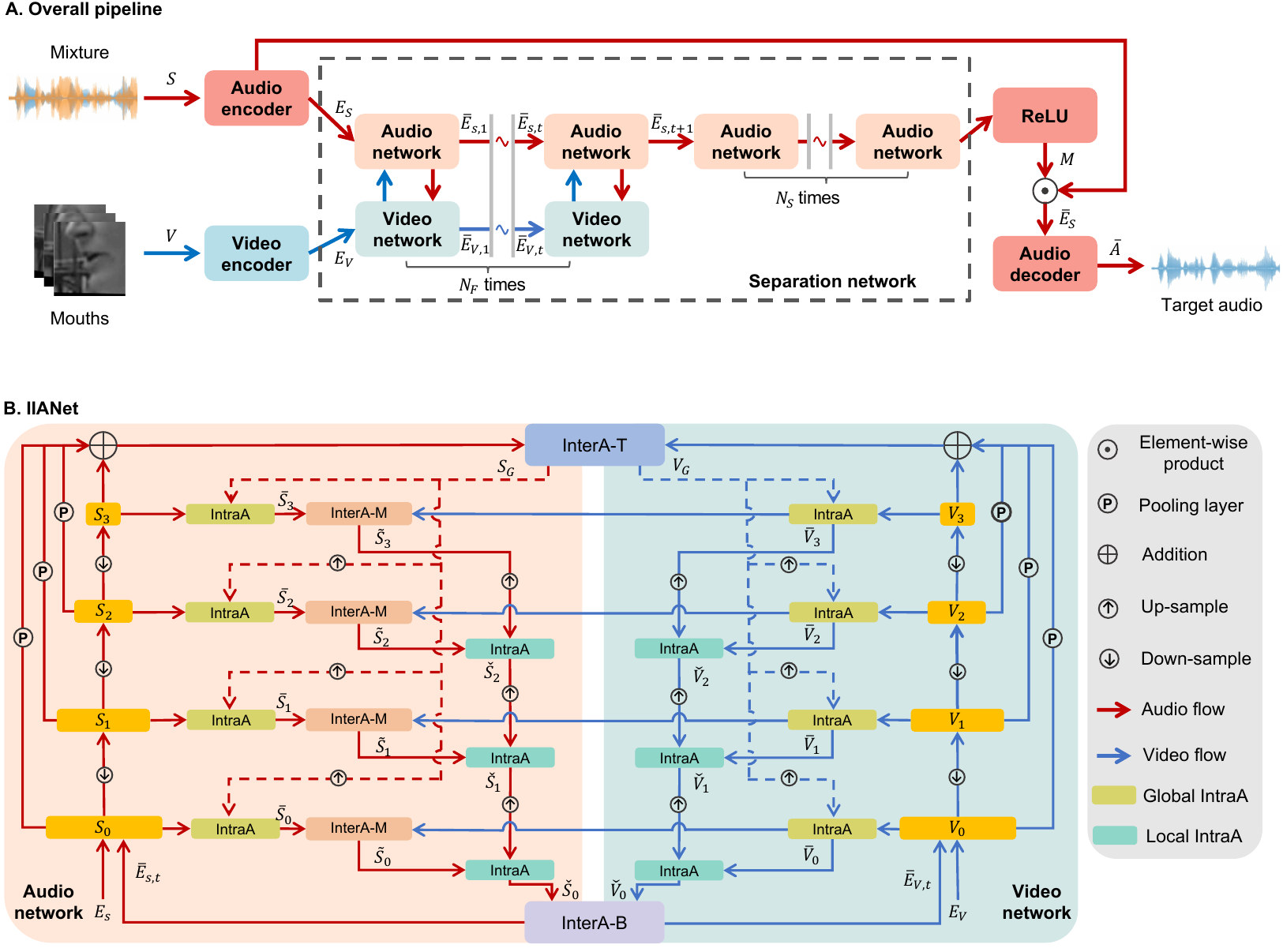}
\caption{The overall pipeline of IIANet. (A) IIANet consists of four main components: audio encoder, video encoder, separation network, and audio decoder. The red and blue tildes indicate that the same module is repeated several times. (B) The separation network contains two types of attention blocks: IntraA and InterA (InterA-T, InterA-M, InterA-B) blocks. 
 The dashed lines indicate the use of global features $\mathbf{S}_G$ and $\mathbf{V}_G$ as top-down attention modulation for multi-scale features $\mathbf{S}_i$ and $\mathbf{V}_i$. 
All blocks use different parameters but keep the same across different cycles.}
\label{fig:all}
\vspace{-15pt}
\end{figure*}

Let $\mathbf{A}\in R^{1\times T_a}$ and $\mathbf{V}\in R^{H\times W\times T_v}$ represent the audio and video streams of an individual speaker, respectively, where $T_a$ denotes the audio length; $H$, $W$, and $T_v$ denote the height, width, and the number of lip frames, respectively. We aim to separate a high-quality, clean single-speaker audio $\mathbf{A}$ from the noisy speech $\mathbf{S}\in R^{1\times T_a}$ based on the video cues $\mathbf{V}$ and filter out the remaining speech components (other interfering speakers). Specifically, our pipeline consists of four modules: audio encoder, video encoder, separation network and audio decoder (see Figure~\ref{fig:all}A).

The overall pipeline of the IIANet is summarized as follows. First, we obtain a video containing two speakers from the streaming media, and the lips are extracted from each image frame. Second, we encode the lip frames and noisy speech into lip embeddings $\mathbf{E}_V\in R^{N_v\times T_v} $ and noisy speech embeddings $\mathbf{E}_S\in R^{N_a\times T'_{a}}$ using a video encoder same as CTCNet \citep{li2022audio} and an audio encoder (a 1D convolutional layer), respectively, where $N_v$, $N_a$ and $T'_{a}$ are the visual and audio embedding dimensions and the number of frames for audio embedding, respectively. Third, the separation network takes $\mathbf{E}_S$ and $\mathbf{E}_V$ as inputs and outputs a soft mask $\mathbf{M}\in R^{N_a\times T'_{a}}$ for the target speaker. We multiply $\mathbf{E}_S$ with $\mathbf{M}$ to estimate the target speaker's speech embedding ($\bar{\mathbf{E}}_S=\mathbf{E}_S\odot \mathbf{M}$), where ``$\odot$" denotes the element-wise product. Finally, we utilize the audio decoder (a 1D transposed convolutional layer) to convert the $\bar{\mathbf{E}}_S$ back to the waveform, generating the target speaker's speech $\bar{\mathbf{A}}\in R^{1\times T_a}$.

\subsection{Audio-visual separation network}

\begin{figure*}[t]
\centering
\includegraphics[width=0.6\linewidth]{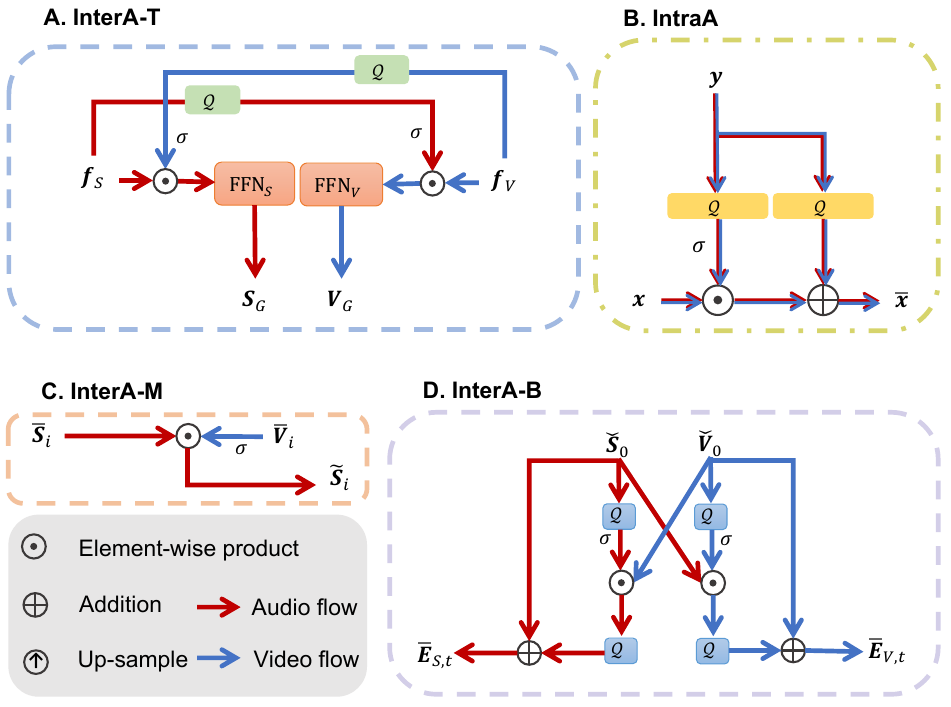}
\caption{Flow diagram of InraA and InterA blocks: (A) InterA-T block, (B) IntraA block, (C) InterA-M block and (D) InterA-B block in the IIANet, where $\odot$ denotes element-wise product and $\sigma$ denotes the sigmoid function. 
}
\label{fig:sablock}
\vspace{-15pt}
\end{figure*}

The core architecture of IIANet is an audio-visual separation network consisting of two types of components: (A) intra-attention (IntraA) blocks, (B) inter-attention (InterA) blocks (distributed at different locations: top (InterA-T), middle (InterA-M) and bottom (InterA-B)). The architecture of the separation network is illustrated in Figure~\ref{fig:all}B, which is described as follows:

\begin{enumerate}
    \item \textbf{Bottom-up pass}. The audio network and video network take $\mathbf{E}_S$ and $\mathbf{E}_V$ as inputs and output multi-scale auditory $\{\mathbf{S}_{i}\in R^{N_a\times \frac{T'_{a}}{2^{i}}}|i=0,1...,D\}$ and visual $\{\mathbf{V}_{i}\in R^{N_v\times \frac{T_{v}}{2^{i}}}|i=0,1...,D\}$ features, where $D$ denotes the total number of convolutional layers with a kernel size of 5 and a stride of 2, followed by a global normalization layer (GLN) \citep{luo2019conv}. In Figure~\ref{fig:all}B, we show an example model with $D=3$.
    
    \item \textbf{AV fusion through the InterA-T block}. The multi-scale auditory $\mathbf{S}_{i}$ and visual $\mathbf{V}_{i}$ features are fused using the InterA-T block (Figure~\ref{fig:sablock}A) at the top of separation network to obtain the inter-modal global features $\mathbf{S}_G\in R^{N_a\times \frac{T'_{a}}{2^D}}$ and $\mathbf{V}_G\in R^{N_v\times \frac{T{v}}{2^D}}$.  
    \item \textbf{AV fusion in the top-down pass through IntraA and InterA-M blocks}. Firstly, the $\mathbf{S}_G$ and $\mathbf{V}_G$ are employed to modulate $\mathbf{S}_{i}$ and $\mathbf{V}_{i}$ within each modality using top-down global IntraA blocks, yielding auditory $\Bar{\mathbf{S}}_{i}\in R^{N_a\times \frac{T'_{a}}{2^i}}$ and visual $\Bar{\mathbf{V}}_{i}\in R^{N_v\times \frac{T_v}{2^i}}$ features. Secondly, the InterA-M block takes the auditory $\Bar{\mathbf{S}}_{i}$ and visual $\Bar{\mathbf{V}}_{i}$ features at the same temporal scale as inputs, and output the modulated auditory features $\tilde{\mathbf{S}}_{i}\in R^{N_a\times \frac{T'_{a}}{2^i}}$. 
    Finally, the top-down pass generates the auditory $ \check{\mathbf{S}}_0 \in R^{N_a\times T'_{a}}$ and visual $\check{\mathbf{V}}_0 \in R^{N_v\times T_{v}}$ features at the maximum temporal scale using top-down local IntraA blocks.
    
    \item \textbf{AV fusion through the InterA-B block}. The InterA-B block (Figure~\ref{fig:sablock}D) takes the auditory $\check{\mathbf{S}}_{0}$ and visual $\check{\mathbf{V}}_{0}$ features as input and outputs audio-visual features $\bar{\mathbf{E}}_{S, t}\in R^{N_a\times T'_{a}}$ and $\bar{\mathbf{E}}_{V, t}\in R^{N_v\times T_{v}}$, where $t$ denotes the cycle index of audio-visual fusion.
    
    
    \item \textbf{Cycle of audio and video networks}. The above steps complete the first cycle of the separation network. For the $t$-th cycle where $t\geq 2$, the audio-visual features $\bar{\mathbf{E}}_{S,t-1}$ and $\bar{\mathbf{E}}_{V,t-1}$ obtained in the $(t-1)$-th cycle, instead of $\mathbf{E}_S$ and $\mathbf{E}_V$, serve as input to the separation network, and we repeat the above steps. After $N_F$ cycles, the auditory features $\bar{\mathbf{E}}_{S, t}$ are refined alone for $N_S$ cycles in the audio network to reconstruct high-quality audio. Finally, we pass the output of the final cycle through an activation function (ReLU) to yield the separation network's output $\mathbf{M}$, as illustrated in Figure~\ref{fig:all}A.
\end{enumerate}

The steps 2-4 and the blocks mentioned in these steps are described in details in what follows.

\noindent{\textbf{Step 2: AV fusion through the InterA-T block}}.
In the InterA-T block, we first utilize an average pooling layer with a pooling ratio $2^{D-i}$ in the temporal dimension to down-sample the temporal dimensions of $\mathbf{S}_{i}$ and $\mathbf{V}_{i}$ to $\frac{T'_{a}}{2^D}$ and $\frac{T_v}{2^D}$, respectively, and then merge them to obtain global features $\mathbf{S}_G$ and $\mathbf{V}_G$ using InterA-T block  (Figure~\ref{fig:sablock}A). The detailed process of the InterA-T block is described by the equations below:
\begin{equation}
\begin{aligned}
\pmb{f}_S&=\sum_{i = 0}^{D-1} p_i(\mathbf{S}_{i}) + \mathbf{S}_{D} \;  \text{and} \; \pmb{f}_V=\sum_{i = 0}^{D-1} p_i(\mathbf{V}_{i}) + \mathbf{V}_{D}, \\ 
\mathbf{S}_{G} &= \text{FFN}_{S}\left(\pmb{f}_S \odot \sigma(\mathcal{Q}(\pmb{f}_V))\right), \\ 
\mathbf{V}_{G} &= \text{FFN}_{V}\left(\pmb{f}_V \odot \sigma(\mathcal{Q}(\pmb{f}_S))\right),
\end{aligned}
\end{equation}
where $\pmb{f}_S\in R^{N_a\times \frac{T'_{a}}{2^D}}$ and $\pmb{f}_V\in R^{N_v\times \frac{T_{v}}{2^D}}$ denote the cumulative multi-scale auditory and visual features, $\mathrm{FFN}(\cdot)$ denotes a feed-forward network with three convolutional layers followed by a GLN, $\mathcal{Q}$\footnote{For the simplification of symbols, we use $\mathcal{Q}$ to denote a 1D convolutional layer followed by a GLN throughout the paper. But unless otherwise specified, different $\mathcal{Q}$'s do not share parameters.} denotes a convolutional layer followed by a GLN, $p_i(\cdot)$ denotes the average pooling layers and $\sigma$ denotes the sigmoid function. Here, $p_i(\cdot)$ uses different pooling ratios for different $i$ such that the temporal dimensions of all $p_i(\mathbf{S}_i)$ are $\frac{T'_{a}}{2^D}$ and the temporal dimensions of all $p_i(\mathbf{V}_i)$ are $\frac{T_v}{2^D}$.  The hyperparameters of $\mathcal{Q}$ are specified such that the resulting audio and video features have the same embedding dimension. The $\sigma$ function receiving information from one modality acts as selective attention, modulating the other modality. The global features $\mathbf{S}_{G}$ and $\mathbf{V}_{G}$ are used to guide the separation process in the top-down pass. 

\noindent{\textbf{Step 3: AV fusion in the top-down pass through IntraA and InterA-M blocks}}.
Given the multi-scale audio-visual features \{$\mathbf{S}_{i}, \mathbf{V}_{i}$\} and $\mathbf{S}_G$ and $\mathbf{V}_G$ as inputs, we employ the global IntraA blocks $\phi(\mathbf{x},\mathbf{y})$ (Figure~\ref{fig:sablock}B) in the top-down pass (the dotted lines in Figure~\ref{fig:all}B) to extract audio $\Bar{\mathbf{S}}_{i}$ and visual features $\Bar{\mathbf{V}}_{i}$ at each temporal scale,
\begin{equation}
    \bar{\mathbf{x}} = \phi(\mathbf{x},\mathbf{y}) 
           = \sigma\left ( \mathcal{Q}\left ( \mu \left ( \mathbf{y} \right )  \right )  \right ) \odot \mathbf{x} + \mathcal{Q}\left ( \mu \left ( \mathbf{y} \right )  \right )
\end{equation}\label{eq:2}
where $\mu(\cdot)$ denotes interpolation up-sampling. For audio signals, $\bar{\mathbf{S}}_i = \phi(\mathbf{S}_i,\mathbf{S}_G)$; for video signals, $\bar{\mathbf{V}}_i = \phi(\mathbf{V}_i,\mathbf{V}_G)$. The hyperparameters of $\mu$ are specified such that the resulted audio and video signals have the same temporal dimension.
Since these attention blocks use the global features $\mathbf{S}_G$ and $\mathbf{V}_G$ to modulate the intermediate features $\mathbf{S}_{i}$ and $\mathbf{V}_{i}$ when reconstructing these features in the top-down process, they are called {\it global} IntraA blocks. This is inspired by the global attention mechanism used in an AOSS model \citep{li2022efficient}. But in that model, the global feature is simply upsampled, then goes through a sigmoid function and multiplied with the intermediate features $\mathbf{x}$. Mathematically, the process is written as follows
\begin{equation}
\bar{\mathbf{x}}=\phi'(\mathbf{x},\mathbf{y})=\sigma(\mu(\mathbf{y}))\odot \mathbf{x}.
\end{equation}\label{eq:3}
We empirically found that $\phi$ worked better than $\phi'$ (see \textbf{Appendix~\ref{sec:global-ia}}).


In contrast to previous AVSS methods \citep{gao2021visualvoice, li2022audio}, we investigate which part of the audio network should be attended to undergo the guidance of visual features within the same temporal scale. We utilize the visual features $\bar{\mathbf{V}}_{i}$ from the same temporal scale as $\bar{\mathbf{S}}_{i}$ to modulate the auditory features $\bar{\mathbf{S}}_{i}$ using the InterA-M blocks (Figure~\ref{fig:sablock}C),
\begin{equation}
\tilde{\mathbf{S}}_{i}= \sigma(\mathcal{Q}(\mu(\bar{\mathbf{V}}_{i})))\odot  \bar{\mathbf{S}}_{i},  \quad  \forall i \in [0, D],
\end{equation}
where $\tilde{\mathbf{S}}_{i}\in R^{N\times \frac{T'_{a}}{2^i}}$ denotes the modulated auditory features. Same as before, the hyperparameters of $\mathcal{Q}$ and $\mu$ are specified such that the two terms on the two sides of $\odot$ have the same dimension. In this way, the InterA-M blocks are expected to extract the target speaker's sound features according to the target speaker's visual features at the same scale. 

Then, we reconstructed auditory and visual features separately to obtain the features $(\check{\mathbf{S}}_{0}, \check{\mathbf{V}}_{0})$ in a top-down pass (Figure~\ref{fig:all}B). We leverage the local top-down attention mechanism (Li et al., 2023), and the corresponding blocks are called {\it local} IntraA blocks, which are able to fuse as many multi-scale features as possible while reducing information redundancy. We start from the $(D-1)$-th layer. The attention signals are $\tilde{\mathbf{S}}_D$ and  $\Bar{\mathbf{V}}_D$. Therefore,
\begin{equation}
     \check{\mathbf{S}}_{D-1}=\phi(\tilde{\mathbf{S}}_{D-1}, \tilde{\mathbf{S}}_{D}), \quad
 \check{\mathbf{V}}_{D-1}=\phi(\bar{\mathbf{V}}_{D-1}, \bar{\mathbf{V}}_{D}).
\end{equation}
For $i=D-2, ...,0$, the attention signals are $\check{\mathbf{S}}_i$ and  $\check{\mathbf{V}}_i$, therefore
\begin{equation}
    \check{\mathbf{S}}_{i}=\phi(\tilde{\mathbf{S}}_{i}, \check{\mathbf{S}}_{i+1}), \quad
 \check{\mathbf{V}}_{i}=\phi(\bar{\mathbf{V}}_{i}, \check{\mathbf{V}}_{i+1}).
\end{equation}
Please note that different IntraA blocks within one cycle, as depicted in Figure~\ref{fig:all}B, use different parameters. 

\noindent{\textbf{Step 4: AV fusion through the InterA-B block}}.
In CTCNet \citep{li2022audio}, the global audio-visual features fused all auditory and visual multi-scale features using the concatenation strategy, leading to high computational complexity. Here, we only perform inter-modal fusion on the finest-grained auditory and visual features $(\check{\mathbf{S}}_{0}$ and $\check{\mathbf{V}}_{0})$ using attention strategy, greatly reducing computational complexity. 

Specifically, we reduce the impact of redundant features by incorporating the InterA-B block (Figure~\ref{fig:sablock}D) between the finest-grained auditory and visual features $\check{\mathbf{S}}_{0}$ and $\check{\mathbf{V}}_{0}$.
In particular, the process of global audio-visual fusion is as follows:
\begin{equation}
\begin{aligned}
\bar{\mathbf{E}}_{S,t}&=\check{\mathbf{S}}_{0}+\mathcal{Q}\left(\mu(\check{\mathbf{V}}_{0}) \odot \sigma\left(\mathcal{Q}\left(\check{\mathbf{S}}_{0}\right)\right)\right),  \\ \bar{\mathbf{E}}_{V,t}&=\check{\mathbf{V}}_{0}+\mathcal{Q}\left(\mu(\check{\mathbf{S}}_{0}) \odot \sigma\left(\mathcal{Q}\left(\check{\mathbf{V}}_{0}\right)\right)\right).
\end{aligned}
\end{equation}
Same as in (1), the hyperparameters of $\mathcal{Q}$ and $\mu$ are specified such that the element-wise product and addition in (5) function correctly. The working process of this block is simple. One modality features are modulated by the other modality features through an attention function $\sigma$, and then the results are added to the original features. Several convolutions and up-samplings are interleaved in the process to ensure that the dimensions of results are appropriate. We keep the original features in the final results by addition because we do not want the final features to be altered too much by the other modality information. 

Finally, the output auditory and visual features $\bar{\mathbf{E}}_{S,t}$ and $\bar{\mathbf{E}}_{V,t}$ serve as inputs for the subsequent audio network and video network.

%% file: experiments.tex
\subsection{Datasets}
Consistent with previous research, we experimented on three commonly used AVSS datasets: LRS2 \cite{afouras2018deep}, LRS3 \cite{afouras2018lrs3}, and VoxCeleb2 \cite{chung2018voxceleb2}. Due to the fact that the LRS2 and VoxCeleb2 datasets were collected from YouTube, they present a more complex acoustic environment compared to the LRS3 dataset. Each audio was 2 seconds in length with a sample rate of 16 kHz. The mixture was created by randomly selecting two different speakers from datasets and mixing their speeches with signal-to-noise ratios between -5 dB and 5 dB. Lip frames were synchronized with audio, having a frame rate of 25 FPS and a size of 88$\times$88 grayscale images. We used the same dataset split consistent with previous works \cite{li2022audio, gao2021visualvoice, lee2021looking}. See \textbf{Appendix~\ref{sec:datasets}} for details.

\begin{table*}[ht]
\footnotesize
\centering
\begin{tabular}{c|ccccccccc}
\toprule
\multirow{2}{*}{Methods} & \multicolumn{3}{c}{LRS2} & \multicolumn{3}{c}{LRS3} & \multicolumn{3}{c}{VoxCeleb2} \\ 
\cmidrule(r){2-4} \cmidrule(r){5-7} \cmidrule(r){8-10}
                       & SI-SNRi       & SDRi  &   PESQ   & SI-SNRi       & SDRi    &  PESQ  & SI-SNRi         & SDRi      & PESQ    \\
                       \midrule
AVConvTasNet  \cite{wu2019time}         & 12.5         & 12.8  &  2.69  & 11.2         & 11.7      &   2.58  & 9.2            & 9.8      &  2.17  \\
LWTNet  \cite{afouras2020self}               & -            & 10.8  &   -   & -            & 4.8    & -  & -              & -    &   -     \\
Visualvoice  \cite{gao2021visualvoice}          & 11.5         & 11.8   & 2.78  & 9.9          & 10.3    &  2.13  & 9.3            & 10.2     &  2.45  \\
CaffNet-C \cite{lee2021looking}             & -            & 10.0   &  1.15   & -            & 9.8    & -  & -              & 7.6    &  -    \\
AVLiT-8 \cite{martel2023audio} & 12.8            & 13.1   &  2.56 & 13.5            & 13.6   & 2.78   & 9.4              & 9.9    &   2.23  \\
CTCNet  \cite{li2022audio}               & 14.3         & 14.6    &  3.08 & 17.4         & 17.5     & 3.24  & 11.9           & 13.1     &  3.00  \\ \midrule
IIANet (\textit{ours})                 & \sethlcolor{myred}\hl{16.0}         & \sethlcolor{myred}\hl{16.2}   &  \sethlcolor{myred}\hl{3.23} & \sethlcolor{myred}\hl{18.3}         & \sethlcolor{myred}\hl{18.5}   &  \sethlcolor{myred}\hl{3.28}  & \sethlcolor{myred}\hl{13.6}           & \sethlcolor{myred}\hl{14.3}     & \sethlcolor{myred}\hl{3.12} \\
IIANet-fast (\textit{ours})  & \sethlcolor{mygreen}\hl{15.1}         & \sethlcolor{mygreen}\hl{15.4}   & \sethlcolor{mygreen}\hl{3.11}  & \sethlcolor{mygreen}\hl{17.8}         & \sethlcolor{mygreen}\hl{17.9}   & \sethlcolor{mygreen}\hl{3.25}  & \sethlcolor{mygreen}\hl{12.6}           & \sethlcolor{mygreen}\hl{13.6}    &  \sethlcolor{mygreen}\hl{3.01}   \\ 
\bottomrule 
\end{tabular}
\caption{Separation results of different AVSS methods on LRS2, LRS3, and VoxCeleb2 datasets. These metrics represent the average values for all speakers in each test set, where larger SI-SNRi, SDRi and PESQ values are better. ``-" denotes results not reported in the original paper. \sethlcolor{myred}\hl{Optimal} and \sethlcolor{mygreen}\hl{suboptimal} performances is highlighted.}
\label{com-all}
\end{table*}

\begin{table*}[ht]
\footnotesize
\centering
\begin{tabular}{c|ccccc}
\toprule
\multirow{2}{*}{Methods}                                                                           & \multicolumn{2}{c}{Computation  Cost} & \multicolumn{2}{c}{Inference  Time} & \multirow{2}{*}{GPU Memory (MB)} \\ \cmidrule{2-3} \cmidrule{4-5}
                                                                                                   & MACs (G)        & Params (M)        & CPU (s)          & GPU (ms)         &                                  \\ \midrule
AVConvTasNet \cite{wu2019time}  & 23.8              & 16.5              & 1.06             & \sethlcolor{mygreen}\hl{62.51}            & 117.05                           \\
Visualvoice \cite{gao2021visualvoice}& \sethlcolor{myred}\hl{9.7}               & 77.8              & 2.98             & 110.31           & 313.74                           \\
AVLiT \cite{martel2023audio}         & 18.2              & \sethlcolor{mygreen}\hl{5.8}              & \sethlcolor{myred}\hl{0.46}             & \sethlcolor{myred}\hl{62.51}            & \sethlcolor{mygreen}\hl{24.0}                             \\
CTCNet \cite{li2022audio}         & 167.1             & 7.0               & 1.26             & 84.17            & 75.8                             \\ \midrule
IIANet (\textit{ours})                                                            & 18.6              & \sethlcolor{myred}\hl{3.1}               & 1.27             & 110.11           & \sethlcolor{myred}\hl{12.5}                             \\ 
IIANet-fast (\textit{ours})                                                      & \sethlcolor{mygreen}\hl{11.9}              & \sethlcolor{myred}\hl{3.1}               & \sethlcolor{mygreen}\hl{0.52}             & 70.94            & \sethlcolor{myred}\hl{12.5}                             \\
\bottomrule
\end{tabular}
\caption{Parameter sizes and computational complexities of different AVSS methods. All results are measured with an input audio length of 1 second, a sample rate of 16 kHz, and a video frame sample rate of 25 FPS. Inference time is measured in the same testing environment without the use of any additional acceleration techniques, such as quantization or pruning.}
\label{effect}\vspace{-15pt}
\end{table*}

\subsection{Model configurations and evaluation metrics}

The IIANet model was implemented using PyTorch \cite{paszke2019pytorch}, and optimized using the Adam optimizer \cite{kingma2014adam} with an initial learning rate of 0.001. The learning rate was cut by 50\% whenever there was no improvement in the best validation loss for 15 consecutive epochs. The training was halted if the best validation loss failed to improve over 30 successive epochs. Gradient clipping was used during training to avert gradient explosion, setting the maximum $L2$ norm to 5. We used the SI-SNR objective function to train the IIANet. See \textbf{Appendix~\ref{sec:loss}} for details. All experiments were conducted with a batch size of 6, utilizing 8 GeForce RTX 4090 GPUs. The hyperparameters are included in \textbf{Appendix~\ref{sec:config}}. 

Same as previous research \cite{gao2021visualvoice, li2022audio, afouras2018deep}, the SDRi \cite{vincent2006performance} and SI-SNRi \cite{le2019sdr} were used as a metric for speech separation. See \textbf{Appendix~\ref{sec:eval}} for details.
To simulate human auditory perception and to assess and quantify the quality of speech signals, we also used PESQ \cite{union2007wideband} as an evaluation metric, which is used to measure the clarity and intelligibility of speech signals, ensuring the reliability of the results.

\subsection{Comparisons with state-of-the-art methods} 
To explore the efficiency of the proposed approach, we have designed a faster version, namely \textit{IIANet-fast}, which runs half of $N_S$ cycles compared to IIANet (see \textbf{Appendix~\ref{sec:config}} for detailed parameters). We compared IIANet and IIANet-fast with existing AVSS methods.
To conduct a fair comparison, we obtained metric values in original papers or reproduced the results using officially released models by the authors.

\textbf{Separation quality}. 
As demonstrated in Table~\ref{com-all}, IIANet consistently outperformed existing methods across all datasets, with at least a 0.9 dB gain in SI-SNRi. In particular, IIANet exceled in complex environments, notably on the LRS2 and VoxCeleb2 datasets, outperforming AVSS methods by a minimum of 1.7 dB SI-SNRi in separation quality. This significant improvement highlights the importance of IIANet's hierarchical audio-visual integration. In addition, IIANet-fast also achieved excellent results, which surpassed CTCNet in separation quality with an improvement of 1.1 dB in SI-SNRi on the challenging LRS2 dataset. 

\textbf{Model size and computational cost}. In practical applications, model size and computational cost are often crucial considerations. We employed three hardware-agnostic metrics to circumvent disparities between different hardware, including number of parameters (Params) and Multiply-Accumulate operations (MACs)\footnote{For applications that need to run models in resource-constrained environments, such as mobile devices and edge computing, MACs is a key metric because it is directly related to power consumption, latency, and storage requirements.}, and GPU memory usage during inference. We also selected two hardware-related metrics to measure inference speed on specific GPU and CPU hardware. The results were shown in Table~\ref{effect}. Please note that some of the AVSS methods \cite{alfouras2018conversation, afouras2020self, lee2021looking} in Table~\ref{com-all} are not included in Table 2 because they are not open sourced. 

Compared to the previous SOTA method CTCNet, IIANet can achieve significantly higher separation quality using 11\% of the MACs, 44\% of the parameters and 16\% of the GPU memory. IIANet's inference on CPU was as fast as CTCNet, but was slower on GPU. With better separation quality than CTCNet, IIANet-fast had much less computational cost (only 7\% of CTCNet’s MACs) and substantially higher inference speed (40\% of CTCNet's time on CPU and 84\% of CTCNet's time on GPU). This not only showcased the efficiency of IIANet-fast but also highlighted its aptness for deployment onto environments with limited hardware resources and strict real-time requirement.

In conclusion, compared to with previous AVSS methods, IIANet achieved a good trade-off between separation quality and computational cost.

\subsection{Multi-speaker performance}

We present results on the LRS2 dataset with 3 and 4 speakers. We created two variants of the dataset for these purposes, named LRS2-3Mix and LRS2-4Mix, both used in our model's training and testing. Correspondingly, the default LRS2 dataset has a new name LRS2-2Mix. Throughout the paper, unless otherwise specified, LRS2 dataset refers to the default LRS2-2Mix dataset.
For more details on dataset construction, training processes, and testing procedures, please refer to \textbf{Appendix \ref{sec:multi-spks}}.

\begin{table*}[h]
\footnotesize
\centering
\begin{tabular}{c|cccccc}
\toprule
\multirow{2}{*}{Methods} & \multicolumn{2}{c}{LRS2-2Mix} & \multicolumn{2}{c}{LRS2-3Mix} & \multicolumn{2}{c}{LRS2-4Mix} \\ 
    \cmidrule(r){2-3} \cmidrule(r){4-5} \cmidrule(r){6-7}                     & SI-SNRi         & SDRi        & SI-SNRi         & SDRi        & SI-SNRi         & SDRi     
                          \\ \midrule
AVConvTasNet \cite{wu2019time}            & 12.5            & 12.8        & 8.2             & 8.8         & 4.1             & 4.6         \\
AVLIT-8   \cite{martel2023audio}               & 12.8            & 13.1        & 9.4             & 9.9         & 5.0             & 5.7         \\
CTCNet  \cite{li2022audio}                 &  \sethlcolor{mygreen}\hl{14.3}           &  \sethlcolor{mygreen}\hl{14.6}        &  \sethlcolor{mygreen}\hl{10.3}            &  \sethlcolor{mygreen}\hl{10.8}        &  \sethlcolor{mygreen}\hl{6.3}             &  \sethlcolor{mygreen}\hl{6.9}         \\
IIANet (\textit{ours})                   &  \sethlcolor{myred}\hl{16.0}            &  \sethlcolor{myred}\hl{16.2}        &  \sethlcolor{myred}\hl{12.6}            &  \sethlcolor{myred}\hl{13.1}        &  \sethlcolor{myred}\hl{7.8}             &  \sethlcolor{myred}\hl{8.3}        \\ \bottomrule
\end{tabular}
\caption{Performance of different models varies with the number of speakers. These results were based on training conducted using the code from published baseline methods.}
\label{tab:mul-spks}
\vspace{-10pt}
\end{table*}

We use the following baseline methods for comparison: AV-ConvTasNet \cite{wu2019time}, AVLIT \cite{martel2023audio}, and CTCNet \cite{li2022audio}. The results are presented in Table \ref{tab:mul-spks}. Each table column shows a different dataset, where the number of speakers in the mixed signal $C$ varies. The models used for evaluating each dataset were specifically trained for separating a corresponding number of speakers. It is evident from the results that the proposed model significantly outperformed the previous methods across all three datasets.

\subsection{Ablation study}
\label{sec:ablation}

To better understand IIANet, we compared each key component using the LRS2 dataset in a completely fair setup, i.e., using the same architecture and hyperparameters in the following experiments.


\textbf{Importance of IntraA and InterA}. 
We investigated the role of IntraA and InterA in model performance. To this end, we constructed a control model (called Control 1) resulted from removing IntraA and InterA blocks from IIANet. Basically, it uses two upside down UNet \cite{ronneberger2015u} architectures as the visual and auditory backbones and fuses visual and auditory features at the finest temporal scale (Figure~\ref{fig:control}A in \textbf{Appendix}). We then added IntraA blocks to Control 1 to obtain a new model, called Control 2 (Figure~\ref{fig:control}B in \textbf{Appendix}). The detailed description of Controls 1 and 2 can be found in \textbf{Appendix~\ref{sec:control}} and the training and inference procedures are the same as IIANet. As shown in Table~\ref{tab:sa-ca}, Control 1 obtained poor results, and adding IntraA blocks improved the results. This improvement validated the effectiveness of the IntraA blocks in the AVSS task, though it was originally proposed for AOSS \cite{li2022efficient}, because the IntraA module obtained intra-modal contextual features. Adding InterA blocks to Control 2 yielded the proposed IIANet, which obtained a substantial separation quality improvement by 2.4 dB and 2.2 dB in SI-SNRi and SDRi metrics, respectively. This proves that the InterA module enables the visual information to exploit and extract the relevant auditory features fully. Taken together, the results show the effectiveness of the IntraA and InterA blocks.

\begin{table}[ht]
\centering
\footnotesize
\begin{tabular}{c|cc}
\toprule
Model      & SI-SNRi/SDRi  & Params (M)/MACs (G) \\ \midrule
Control 1                    & 13.1/13.4 & \sethlcolor{myred}\hl{2.2/16.7}       \\
Control 2 & \sethlcolor{mygreen}\hl{13.6/14.0} & \sethlcolor{mygreen}\hl{2.2/18.1}      \\ 
IIANet & \sethlcolor{myred}\hl{16.0/16.2} & 3.1/18.6      \\ 
\bottomrule
\end{tabular}
\caption{Importance of IntraA blocks and InterA blocks on the LRS2 test set. MACs are measured with 1s input audio sampled at 16 kHz.}
\label{tab:sa-ca}
\end{table}

\begin{table}[ht]
\centering
\footnotesize
\begin{tabular}{ccc|cc}
\toprule
InterA-T & InterA-M & InterA-B & SI-SNRi / SNRi \\ \midrule
$\surd$   & $\times$   & $\times$   & 13.8   / 14.1 \\
$\times$   & $\surd$   & $\times$   & 14.5   / 14.7 \\
$\times$   & $\times$   & $\surd$   & 14.1   / 14.4 \\ \midrule
$\surd$   & $\surd$   & $\times$   & 14.8   / 15.1 \\
$\surd$   & $\times$   & $\surd$   & 15.0   / 15.2 \\
$\times$   & $\surd$   & $\surd$   & \sethlcolor{mygreen}\hl{15.5   / 15.7} \\ \midrule
$\surd$   & $\surd$   & $\surd$   & \sethlcolor{myred}\hl{16.0   / 16.2}\\ \bottomrule
\end{tabular}
\caption{Impact of each InterA block in IIANet on the LRS2 test set. 
}
\label{ablation}\vspace{-10pt}
\end{table}

\textbf{Importance of different InterA blocks}. Table~\ref{ablation} reports the performances of different combinations of the three InterA blocks: InterA-T, InterA-M and InterA-B. Please note that excluding InterA-T block means directly using $\sum_{i=0}^{D-1} p(\mathbf{S}_i)+\mathbf{S}_D$ and $\sum_{i=0}^{D-1} p(\mathbf{V}_i)+\mathbf{V}_D$ as input to auditory and visual MLPs, respectively, with cross-modal attention signal removed. Excluding InterA-M block means removing visual fusion branches at each temporal scale. Excluding InterA-B block means not performing fine-grained audio-visual fusion.

When only one type of attention blocks is enabled in the inter-modal settings, the InterA-M block results in the highest separation quality. 
When two types of attention blocks are enabled, the combination of InterA-M and InterA-B yielded the best results. The combination of all three types of blocks performed the best. 


\textbf{More analyses}. We also performed additional analyses of IIANet. First, we found that networks that include visual information significantly improve speech separation performance compared to networks that rely only on audio information (see \textbf{Appendix~\ref{sec:audionet}} for details). Then, we improve the separation quality by gradually increasing the number of fusions (1 to 5) (\textbf{Appendix~\ref{sec:av-fusion}}). Meanwhile, we investigated the separation quality of three different pre-trained video encoder models on the LRS2-2Mix dataset (\textbf{Appendix~\ref{sec:vencoder}}). 
In addition, we explored the relationship between the loss of visual cues (especially the speaker's side profile) and the separation quality (\textbf{Appendix~\ref{sec:cues}}). The samples from the side view angle showed a significant decrease in separation quality compared to the frontal view. Compared to other AVSS methods, IIANet had the smallest decrease of about 4\%. Finally, we applied a dynamic mixing data enhancement scheme and obtained even better results
(\textbf{Appendix~\ref{sec:dm}}). Unless specifically stated, experiments did not utilize dynamic mixing data enhancement. 

\subsection{Qualitative evaluation}
We visually compare the speech separation results of AV-ConvTasNet, Visualvoice, CTCNet, and our IIANet, all trained on the LRS2 dataset. We use the spectrogram of the separated audios to demonstrate the quality of the separated signals, as shown in Figure~\ref{fig:spec} in \textbf{Appendix}. In the left part, we observe that IIANet achieves better reconstruction results. See \textbf{Appendix~\ref{sec:visual}} for details.

Besides, we evaluated different AVSS methods (Visualvoice, CTCNet) in real-world scenarios by collecting multi-speaker videos on YouTube. Some typical results are presented in website\footnote{\url{https://anonymous.4open.science/w/IIANet-1251}}. By listening to these results, one can confirm that IIANet generated higher-quality separated audio than other separation models.

%% file: conclusion.tex
We proposed a brain-inspired AVSS model, which is characterized by extensive use of intra-attention and inter-attention in the audio and video networks.
Compared with the previous SOTA method CTCNet on three benchmark datasets, IIANet achieved significantly higher separation quality with 11\% MACs, 44\% parameters and 16\% GPU memory. By reducing the number of cycles, IIANet ran much faster than CTCNet yet still obtained better separation results. 


%% file: appendix.tex
\section{Evaluation on different global IntraA implementations}
\label{sec:global-ia}

In this experiment, we investigated the performance of two implementations of global IntraA blocks in IIANet: $\phi(\mathbf{x},\mathbf{y})$ and $\phi'(\mathbf{x},\mathbf{y})$ (see eqns. (2) and (3)). T
The results in Table 6 indicate that global IntraA blocks with $\phi(\mathbf{x},\mathbf{y})$ significantly improves separation quality without notably increasing computational complexity. This finding underscores the importance of global IntraA with $\phi(\mathbf{x},\mathbf{y})$ in effectively and accurately handing intra-modal contextual features, thereby substantially improving performance.

\begin{table}[ht]
\centering
\begin{tabular}{c|cccc}
\toprule
Implementations      & SI-SNRi & SDRi  & Params (M) &MACs (G) \\ \midrule
$\phi(\mathbf{x},\mathbf{y})$     & \sethlcolor{myred}\hl{16.0} & \sethlcolor{myred}\hl{16.2} & \sethlcolor{myred}\hl{3.1} & \sethlcolor{mygreen}\hl{18.64}       \\
$\phi'(\mathbf{x},\mathbf{y})$ & \sethlcolor{mygreen}\hl{14.7} & \sethlcolor{mygreen}\hl{15.0} & \sethlcolor{myred}\hl{3.1} & \sethlcolor{myred}\hl{17.53}      \\ 
\bottomrule
\end{tabular}
\caption{Comparison of separation quality and efficiency among different global IntraA implementations. 
}
\label{tab:sa}
\end{table}

\section{Dataset details}
\label{sec:datasets}

\textbf{Lip Reading Sentences 2 (LRS2)} \cite{afouras2018deep}. The LRS2 dataset consists of thousands of BBC video clips, divided into Train, Validation, and Test folders. We used the same dataset consistent with previous works \cite{li2022audio, gao2021visualvoice, lee2021looking}, created by randomly selecting two different speakers from LRS2 and mixing their speeches with signal-to-noise ratios between -5dB and 5dB. Since the LRS2 data contains reverberation and noise, and the overlap rate is not 100\%, the dataset is closer to real-world scenarios. We use the same data split containing 11-hour training, 3-hour validation, and 1.5-hour test sets.


\textbf{Lip Reading Sentences 3 (LRS3)} \cite{afouras2018lrs3}. The LRS3 dataset includes thousands of spoken sentences from TED and TEDx videos, divided into Trainval and Test folders. We used the same dataset consistent with previous works \cite{li2022audio, gao2021visualvoice, lee2021looking}, which is constructed by randomly selecting voices of two different speakers from the LRS3 data. In contrast to LRS2, the LRS3 data has relatively less noise and is closer to separation tasks in clean environments. It comprises 28-hour training, 3-hour validation, and 1.5-hour test sets.


\textbf{VoxCeleb2} \cite{chung2018voxceleb2}. The VoxCeleb2 dataset contains over one million sentences from 6,112 individuals extracted from YouTube videos, divided into Dev and Test folders. We used the same dataset consistent with previous works \cite{li2022audio, gao2021visualvoice, lee2021looking}, constructed by selecting 5\% of the data from the Dev folder of VoxCeleb2 for creating training and validation sets. Similar to LRS2, VoxCeleb2 also contains a significant amount of noise and reverberation, making it closer to real-world scenarios, but the acoustic environment of VoxCeleb2 is more complex and challenging. It comprises 56-hour training, 3-hour validation, and 1.5-hour test sets.

\section{Objective function}\label{sec:loss}
We used scale-invariant source-to-noise ratio (SI-SNR) \cite{le2019sdr} as the objective function to be maximized for training IIANet. The SI-SNR for each speaker is defined as:
\begin{equation}
\text{SI-SNR}(\mathbf{A}, \Bar{\mathbf{A}}) = 10\log_{10} \left(\frac{||\pmb{\omega} \times \mathbf{A}||^2}{||\Bar{\mathbf{A}} - \pmb{\omega} \times \mathbf{A}||^2}\right), \text{ where } \pmb{\omega} =
\frac{\Bar{\mathbf{A}}^\top \mathbf{A}}{\mathbf{A}^\top\mathbf{A}},
\end{equation}
where $\mathbf{A}$ and $\Bar{\mathbf{A}}$ denote the ground truth speech signal and the estimate speech signal by the model, respectively. ``$\times$" denotes the matrix multiplication.

\section{Model configurations}\label{sec:config}

We set the kernel size of the audio encoder and decoder to 16 and stride size to 8. The video encoder utilized a lip-reading pre-trained model consistent with CTCNet \cite{li2022audio} to extract visual embeddings. The number of down-sampling $D$ for auditory and visual networks was set to 4, and the number of channels for all convolutional layers was set to 512. The auditory and visual networks used the same set of weights across different cycles, and they were cycled $N_F=4$ times; after that, the audio network was additionally cycled $N_S=12$ times. The fast version of IIANet, namely IIANet-fast, performed only $N_S=6$ cycles on the auditory network. For FFN in Inter-T blocks, we set the three convolutional layers to have channel size $(512, 1024, 512)$, kernel size $(1, 5, 1)$, stride size $(1, 1, 1)$, and biases (False, True, False). To avoid overfitting, we set the dropout probability for all layers to 0.1.

\begin{figure}[h]
\centering
\includegraphics[width=\linewidth]{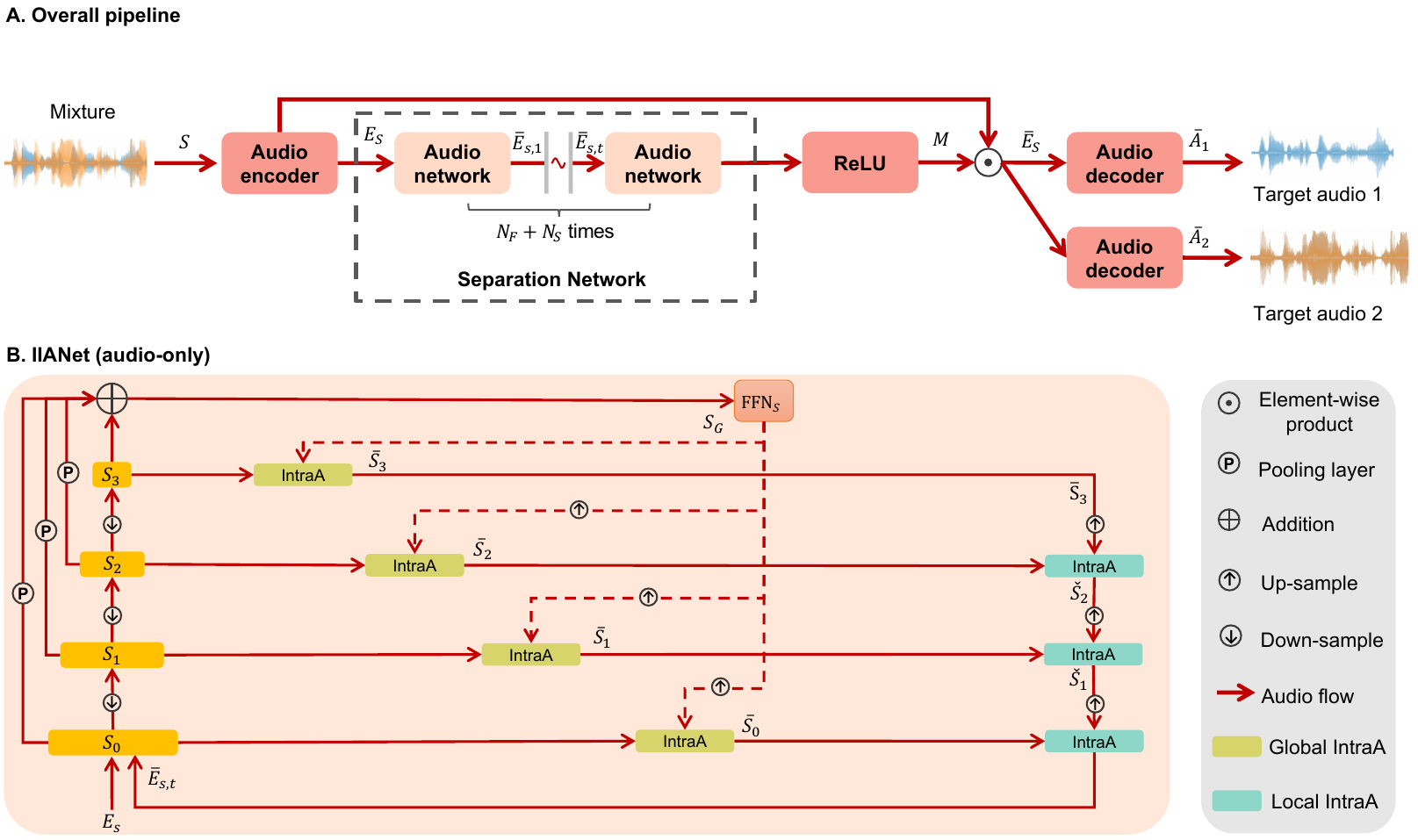}
\caption{The overall pipeline and architecture of IIANet (audio-only).}
\label{fig:audio-net}
\end{figure}

\section{Evaluation metrics}\label{sec:eval}
The quality of speech separation is usually assessed using two key metrics: scale-invariant signal-to-noise ratio improvement (SI-SNRi) and signal distortion ratio improvement (SDRi). The use of SI-SNR and SDR improvesment, which means that it considers the quality of the separated speech and the difficulty of the original blend. This is important because for more difficult speech blends to separate, even if the performance of the speech separation system is not optimal, the improvement relative to the original blend may still be significant. The values of both metrics are derived from the scale-invariant signal-to-noise ratio (SI-SNR) \cite{le2019sdr} and the source distortion ratio (SDR) \cite{vincent2006_SDR}. The higher the values of these metrics, the higher the speech separation quality. The formulas for SI-SNRi, SDR and SDRi are as follows:
\begin{equation}
    \text{SI-SNRi}(\mathbf{S},\mathbf{A}, \Bar{\mathbf{A}}) = \text{SI-SNR}(\mathbf{A}, \Bar{\mathbf{A}}) - \text{SI-SNR}(\mathbf{A},\mathbf{S}),
\end{equation}
\vspace{-13pt}
\begin{equation}
\begin{split}
\text{SDR}(\mathbf{A}, \Bar{\mathbf{A}}) = 10\log_{10} \left( \frac{||\mathbf{A}||^{2}}{||\mathbf{A} - \Bar{\mathbf{A}}||^{2}} \right), \\
\text{SDRi}(\mathbf{S},\mathbf{A}, \Bar{\mathbf{A}}) = \text{SDR}(\mathbf{A}, \Bar{\mathbf{A}}) - \text{SDR}(\mathbf{A},\mathbf{S}),
\end{split}
\end{equation}
where $\mathbf{S}$ denotes the mixture audio.

\section{Audio network}
\label{sec:audionet}

In IIANet (audio-only), we use only audio as input and its pipeline is shown in Figure~\ref{fig:audio-net}A. The structure of the audio network is illustrated in Figure~\ref{fig:audio-net}B. Specifically, the audio network takes $\mathbf{E}_{S,t+1}$ as input and obtains multi-scale auditory features $\mathbf{S}_i$ through a bottom-up pathway. At the coarsest temporal scale, these multi-scale features are integrated to obtain the global feature $\mathbf{S}_G$. Subsequently, we employ a top-down attention mechanism to generate modulated multi-scale auditory features $\bar{\mathbf{S}}_i$, enabling auditory focus across different temporal scales and capturing the core semantics at various scales. Finally, through top-down processing, refined auditory features $\bar{\mathbf{E}}_{S, t}=\check{\mathbf{S}}_{0}$ are produced, serving as the input for the next audio network iteration. To the end, we achieve optimization of auditory features within the audio network with shared parameters.

For the training and testing of IIANet (audio-only), we used hyperparameter settings and datasets consistent with IIANet for a fair comparison with other models. We solved the permutation problem \cite{hershey2016deep} using permutation invariant training (PIT) method \cite{yu2017permutation} consistent with blind source separation methods \cite{li2022efficient, luo2019conv, luo2020dual} to maximize the SI-SNR. 

As shown in Table~\ref{tab:audio-only}, the speech separation quality with added visual information (IIANet) was improved by about 4 dB compared to IIANet (audio-only).

\begin{table*}[ht]
\small
\centering
\begin{tabular}{c|ccccccccc}
\toprule
\multirow{2}{*}{Model} & \multicolumn{3}{c}{LRS2} & \multicolumn{3}{c}{LRS3} & \multicolumn{3}{c}{VoxCeleb2} \\ 
\cmidrule(r){2-4} \cmidrule(r){5-7} \cmidrule(r){8-10}
                       & SI-SNRi       & SDRi  &   PESQ   & SI-SNRi       & SDRi    &  PESQ  & SI-SNRi         & SDRi      & PESQ    \\
                       \midrule
IIANet (\textit{ours})                 & \sethlcolor{myred}\hl{16.0}         & \sethlcolor{myred}\hl{16.2}   &  \sethlcolor{myred}\hl{3.23} & \sethlcolor{myred}\hl{18.3}         & \sethlcolor{myred}\hl{18.5}   &  \sethlcolor{myred}\hl{3.28}  & \sethlcolor{myred}\hl{13.6}           & \sethlcolor{myred}\hl{14.3}     & \sethlcolor{myred}\hl{3.12} \\
IIANet (audio-only) & \sethlcolor{mygreen}\hl{11.2}         & \sethlcolor{mygreen}\hl{11.5}  &  \sethlcolor{mygreen}\hl{2.32}  & \sethlcolor{mygreen}\hl{12.5}        & \sethlcolor{mygreen}\hl{12.8}      &   \sethlcolor{mygreen}\hl{2.54}  & \sethlcolor{mygreen}\hl{9.5}            & \sethlcolor{mygreen}\hl{10.0}      &  \sethlcolor{mygreen}\hl{2.23}  \\
\bottomrule 
\end{tabular}
\caption{Separation results of different AVSS methods on LRS2, LRS3, and VoxCeleb2 datasets. These metrics represent the average values for all speakers in each test set. The larger SI-SNRi, SDRi, and PESQ values are better.}
\label{tab:audio-only}
\end{table*}

\begin{figure}[h]
\centering
\includegraphics[width=0.9\linewidth]{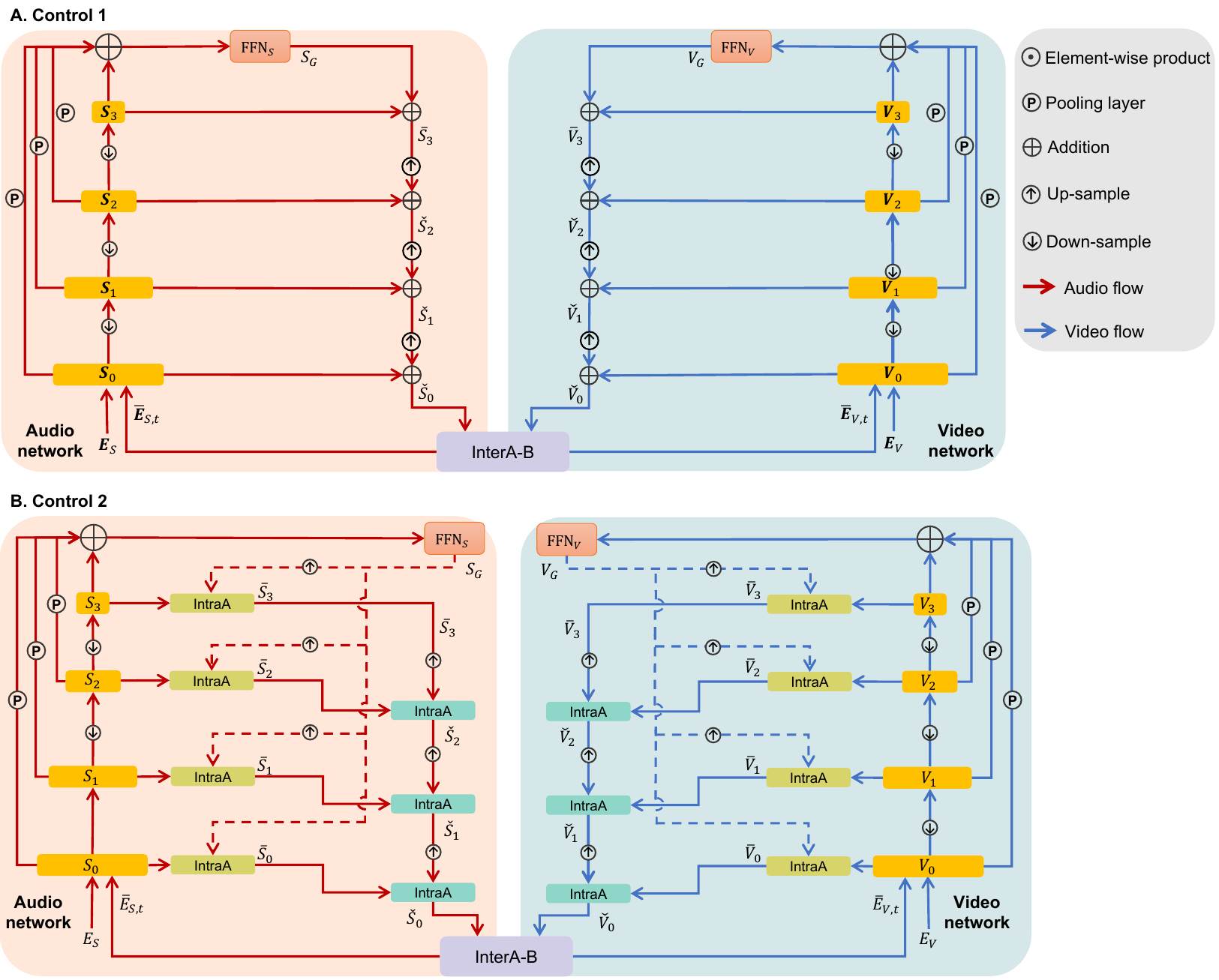}
\caption{The architecture of IIANet's control models. (A) Control 1. It is obtained by removing the IntraA and InterA blocks of IIANet. (B) Control 2. It is obtained by removing InterA blocks of IIANet.}
\label{fig:control}
\end{figure}

\section{Experimental configurations for multiple speakers}\label{sec:multi-spks}
\textbf{Dataset}: We utilized the LRS2-3Mix and LRS2-4Mix datasets for training purposes. For audio data construction, we selected random audio clips from 3 to 4 random speakers mixed with a random signal-to-noise ratio (SNR) value ranging between -5 and 5 dB. Regarding visual data construction, we employed the same experimental setup used in the LRS2-2Mix dataset. In the end, the constructed LRS2-3Mix and LRS2-4Mix datasets comprised 20,000 mixed speech samples for the training set, 5,000 for the validation set, and 300 for the testing set.

\textbf{Training Process}: Consistent with the experimental configuration used for training with two speakers, we trained different baseline models and the IIANet on multi-speaker datasets. This approach allowed us to maintain uniformity in our experimental configuration while exploring the model's performance across varying numbers of speakers.

\textbf{Inference Process}: In the case of video with multiple speakers, our inference process is as follows: Firstly, we use a face detection model \cite{zhang2016joint} to extract the facial images of the different speakers from the video. These images are then cropped to isolate the lip area, with each of these cropped images serving as inputs to the lip reading model. At the same time, the audio corresponding to the video is fed as input into the audio encoder. These processed visual and auditory inputs are subsequently fed into our visual and auditory networks in the separation network, generating visual and auditory features. Through this separation network, we obtain the audio mask of the intended speaker. Then, by employing the audio decoder, we obtain the waveform of the intended speaker. This process continues iteratively until all speakers' visual information has been processed.

\section{Control models}
\label{sec:control}


To validate the effectiveness of our proposed intra-attention and inter-attention blocks in IIANet, we constructed two control models: Control 1 and Control 2. 

Control 1 was obtained by removing all IntraA and InterA blocks from IIANet, and its architecture is depicted in Figure~\ref{fig:control}A. We retained the InterA-B blocks in Control 1 to preserve essential connectivity between the audio and visual networks for audio-visual feature fusion. Control 2 was obtained by adding IntraA blocks to Control 1, and its architecture is depicted in Figure~\ref{fig:control}B. 

\section{Impact of different AV fusion cycles $N_F$}\label{sec:av-fusion}
In IIANet, we gradually expanded the number of AV fusion cycles. With fixed audio-only cycle numbers of 14, the cycle numbers for AV fusion were 1, 2, 3, 4, and 5. This allowed us to gradually detect the effect of visual information through the strength of the fused information. Table ~\ref{tab:nf} shows that separation quality improved as the number of fusions increased, but reached a plateau at 4. 
To make the model lightweight without compromising separation quality, we consider $N_F=4$ to be a good choice.

\begin{table}[ht]
\footnotesize
\centering
\begin{tabular}{c|ccccc}
\toprule
$N_F$ & SI-SNRi & SDRi & PESQ & Params(M) & MACs(G) \\
\midrule
1        & 14.0 & 14.2 & 3.06 & \sethlcolor{myred}\hl{3.1} & \sethlcolor{myred}\hl{17.99}\\
2        & 14.5 & 14.7 & 3.10 & \sethlcolor{myred}\hl{3.1} & \sethlcolor{mygreen}\hl{18.57}\\
3        & \sethlcolor{mygreen}\hl{15.2} & 15.3 & \sethlcolor{mygreen}\hl{3.15} & 3.1 & 18.61\\
4        & \sethlcolor{myred}\hl{16.0} & \sethlcolor{mygreen}\hl{16.2} & \sethlcolor{myred}\hl{3.23} & 3.1 & 18.64 \\
5        & \sethlcolor{myred}\hl{16.0} & \sethlcolor{myred}\hl{16.3} & \sethlcolor{myred}\hl{3.23} & 3.1 & 18.68\\
\bottomrule
\end{tabular}
\caption{Results on the LRS2 dataset for different numbers of audio-visual fusion cycles used by IIANet.}
\label{tab:nf}
\end{table}

\section{Comparison of different video encoders}\label{sec:vencoder}
We examined the separation quality of three different pre-trained video encoder models on the LRS2-2Mix dataset: DC-TCN~\cite{ma2021lip}, MS-TCN~\cite{martinez2020lipreading}, and CTCNet-Lip~\cite{li2022audio}, where DC-TCN, MS-TCN and CTCNet-Lip are lip-reading models that are used to compute embeddings in linguistic contexts. We trained IIANet by replacing only the video encoder. The results are shown in Table~\ref{tab:video-encoder}. Interestingly, we can find that the separation quality was as good as using different video encoders (with SI-SNRi less than 0.3 dB). This indicates that different video encoders do not affect overall performance as much, implying the pivotal role of the audio-video fusion strategy itself. 

\begin{table}[ht]
\centering
\begin{tabular}{c|cccc}
\toprule
Video encoder & SI-SNRi & SDRi & PESQ & Acc(\%) \\
\midrule
DC-TCN        &     15.7    &   16.0   &  3.19    &  88.0\\
MS-TCN        &     \sethlcolor{mygreen}\hl{15.8}    &   \sethlcolor{mygreen}\hl{16.0}   &  \sethlcolor{mygreen}\hl{3.22}    &  \sethlcolor{myred}\hl{85.3} \\
CTCNet-Lip    &     \sethlcolor{myred}\hl{16.0}    &   \sethlcolor{myred}\hl{16.2}   &  \sethlcolor{myred}\hl{3.23}    &  \sethlcolor{mygreen}\hl{84.1}\\
\bottomrule
\end{tabular}
\caption{Results of different video encoders used in IIANet on the LRS2 dataset. ``Acc" denotes the accuracy of lip-reading recognition, which is an effective metric for evaluating lip-reading capabilities.}
\label{tab:video-encoder}
\end{table}

\section{Impaired visual cues}\label{sec:cues}
\begin{figure*}[h]
\centering
\includegraphics[width=1.0\linewidth]{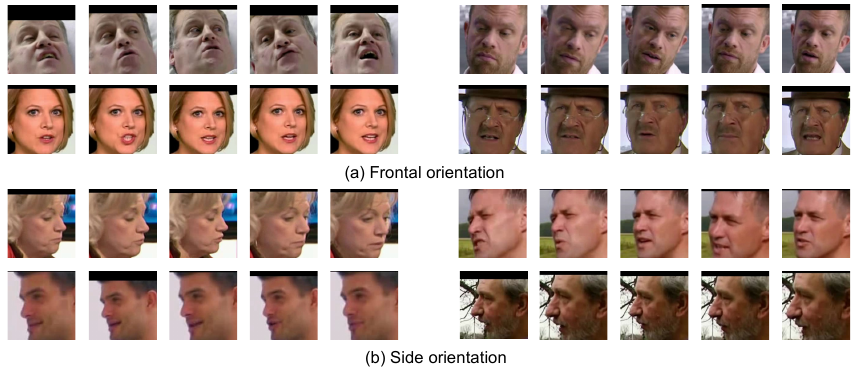}
\caption{Sample visualizations of faces with different orientations in the LRS2 dataset.}
\label{fig:face}
\end{figure*}

We explored the relationship between visual cue loss (speaker's facial orientation, specifically the side profile) and separation quality. We utilized the MediaPipe\footnote{https://github.com/google/mediapipe} tool on the LRS2 dataset to assess the speaker's facial orientation and categorized 6000 audio-video pairs into two groups: frontal orientation (5331 samples) and side orientation (669 samples). A few sample visualizations are shown in Figure~\ref{fig:face}.
Our findings revealed a marked decline in separation quality for samples with a side orientation compared to those facing front. 
Specifically, we observe that the SI-SNRi value decreases by more than 8\% on AVLiT-8 and CTCNet for side orientation samples compared to frontal orientation samples, whereas this decrease is only about 4\% on IIANet. This result underscores the effectiveness of our proposed cross-modal fusion approach in integrating visual and auditory features capable of mitigating visual information impairment.

\begin{table}[ht]
\centering
\begin{tabular}{c|cccc}
\toprule
Model & SI-SNRi & SDRi & PESQ\\
\midrule
\multicolumn{4}{c}{Frontal orientation} \\
\midrule
AVLiT-8 & 13.5 & 13.8 & 2.75 \\
CTCNet  & \sethlcolor{mygreen}\hl{14.9} & \sethlcolor{mygreen}\hl{15.1} & \sethlcolor{mygreen}\hl{3.12} \\
IIANet  & \sethlcolor{myred}\hl{16.3} & \sethlcolor{myred}\hl{16.5} & \sethlcolor{myred}\hl{3.24} \\
\midrule
\multicolumn{4}{c}{Side orientation} \\
\midrule
AVLiT-8 & 12.1 & 12.4 & 2.37 \\
CTCNet  & \sethlcolor{mygreen}\hl{13.7} & \sethlcolor{mygreen}\hl{14.1} & \sethlcolor{mygreen}\hl{3.04} \\
IIANet  & \sethlcolor{myred}\hl{15.7} & \sethlcolor{myred}\hl{15.9} & \sethlcolor{myred}\hl{3.22} \\
\bottomrule
\end{tabular}
\caption{Results of different models with different orientations on the LRS2 dataset.}
\label{tab:visual-cues}
\end{table}

\section{Data augmentation}\label{sec:dm}
To enhance the model's generalization capability, we apply a dynamic mixture data augmentation scheme to the audio-visual separation task. Specifically, we construct new training data during the training process by randomly sampling two speech signals and mixing them after random gain adjustments. Similar approaches have been used in speech and music separation methods \cite{subakan2021attention,zeghidour2021wavesplit}. As shown in Table~\ref{tab:dm}, adopting the dynamic mixing training scheme can further improve the model's separation performance. We recommend using this training scheme in practice, as the cost of generating mixtures is not high (compared to fixed training data, it only adds 0.1 minutes per epoch).
\begin{table*}[ht]
\footnotesize
\centering
\begin{tabular}{c|cccccccccc}
\toprule
\multirow{2}{*}{Method} & \multicolumn{3}{c}{LRS2} & \multicolumn{3}{c}{LRS3} & \multicolumn{3}{c}{VoxCeleb2} & \multirow{2}{*}{Training Time (m)}\\ 
\cmidrule(r){2-4} \cmidrule(r){5-7} \cmidrule(r){8-10}
                       & SI-SNRi       & SDRi  &   PESQ   & SI-SNRi       & SDRi    &  PESQ  & SI-SNRi         & SDRi      & PESQ    \\
                       \midrule
Without DM & \sethlcolor{mygreen}\hl{16.0} & \sethlcolor{mygreen}\hl{16.2} & \sethlcolor{mygreen}\hl{3.23} & \sethlcolor{mygreen}\hl{18.3} & \sethlcolor{mygreen}\hl{18.5} & \sethlcolor{mygreen}\hl{3.28} & \sethlcolor{mygreen}\hl{13.6} & \sethlcolor{mygreen}\hl{14.3} & \sethlcolor{mygreen}\hl{3.12} & \sethlcolor{myred}\hl{36.1}\\
With DM    & \sethlcolor{myred}\hl{16.8} & \sethlcolor{myred}\hl{17.0} & \sethlcolor{myred}\hl{3.35} & \sethlcolor{myred}\hl{18.6} & \sethlcolor{myred}\hl{18.8} & \sethlcolor{myred}\hl{3.30} & \sethlcolor{myred}\hl{14.0} & \sethlcolor{myred}\hl{14.7} & \sethlcolor{myred}\hl{3.17} & \sethlcolor{mygreen}\hl{36.2} \\
\bottomrule 
\end{tabular}
\caption{Comparison of separation performance for different training schemes. "DM" stands for dynamic mixture scheme. Training time refers to the time spent training one epoch on the LRS2 training set.}
\label{tab:dm}
\end{table*}

\section{Visual results}
\label{sec:visual}
\begin{figure*}[t]
\centering
\includegraphics[width=1.0\linewidth]{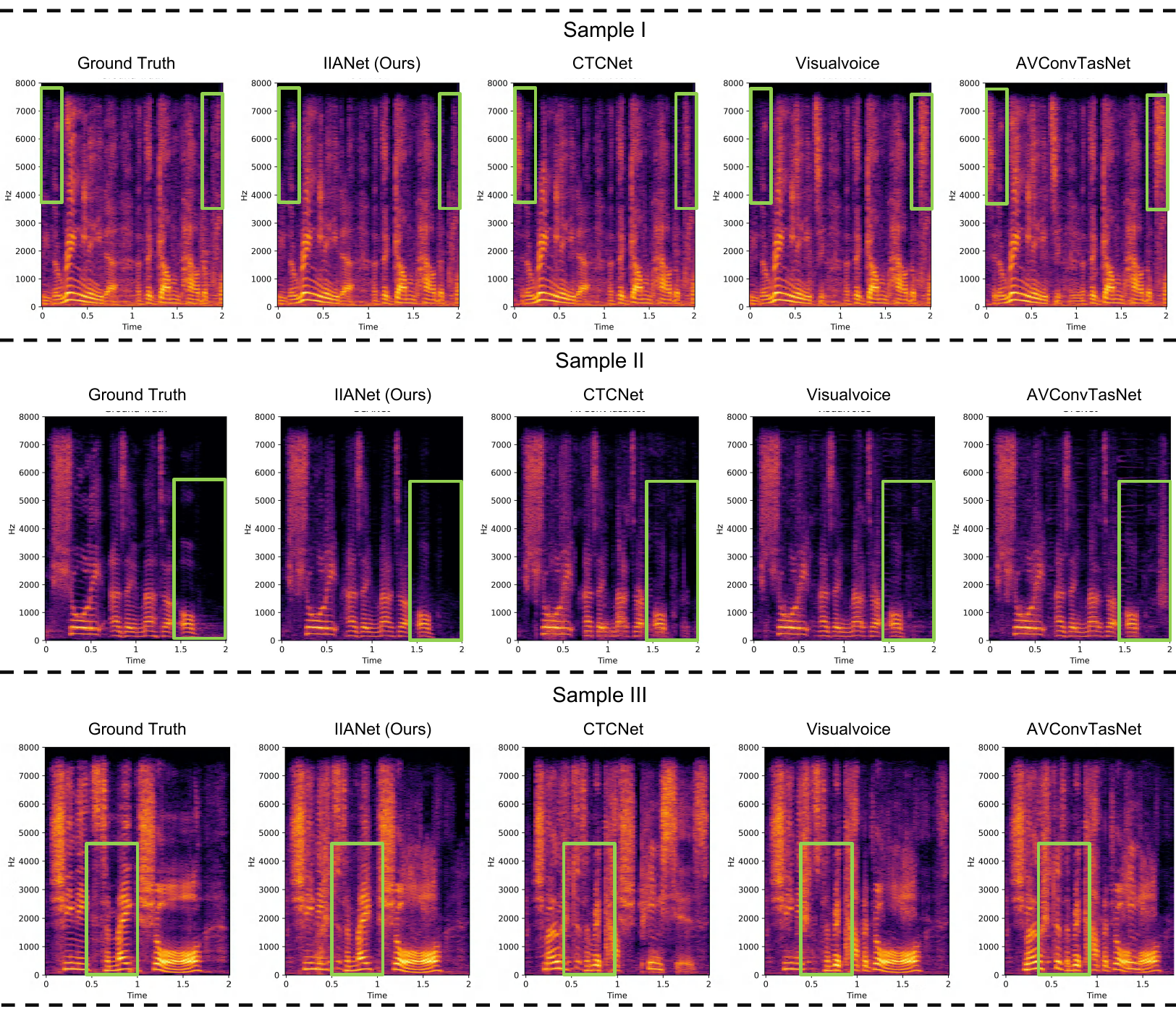}
\caption{Spectrogram of separated audio from different models. Each row represents the results for a same audio mixture.}
\label{fig:spec}
\end{figure*}

We performed visual analysis of the separation performance of four AVSS methods: IIANet, CTCNet, Visualvoice and AVConvTasNet. The results, based on models trained on the LRS2 dataset, are presented in Figure~\ref{fig:spec}. We tested the performance of the four algorithms by randomly selecting three audio mixing samples from the LRS2 dataset.

In Sample I, poor separation of high-frequency components was observed for CTCNet, Visualvoice, and AVConvTasNet. In contrast, the IIANet model exhibited higher separation quality, evidenced by its clarity and strong similarity to the ground truth.	

In Sample II, IIANet presented more detailed harmonic features by accurately removing low and mid-frequency components to the right of the center. In sharp contrast, the other models demonstrated significant deficiencies in recovering harmonic features, resulting in large deviations from the ground truth.	

In Sample III, IIANet demonstrated excellent ability to maintain the complex harmonic structure of the original audio. On the other hand, models such as CTCNet and AVConvTasNet showed blurring effects in this regard, failing to capture the fine details in the ground truth.

